\documentclass[
reprint,
amsmath,amssymb,
aps,
pra,
]{revtex4-1}
\usepackage{float}
\usepackage{subfig}
\usepackage[utf8]{inputenc}
\usepackage{graphicx}% Include figure files
\usepackage{dcolumn}% Align table columns on decimal point
\usepackage{bm}% bold math
\usepackage{hyperref}% add hypertext capabilities
\begin{document}

\preprint{APS/123-QED}

\title{Generalized q-plates and novel vector beams}% Force line breaks with \\

\author{Martin Vergara}
\email{marto@df.uba.ar}
\author{Claudio Iemmi}%
\affiliation{Laboratorio de Procesado de Imágenes, Departamento de Física, Facultad de Ciencias Exactas y Naturales, Universidad de Buenos Aires}%
\affiliation{Consejo Nacional de Investigaciones Científicas y Técnicas}

\date{\today}% It is always \today, today,
\begin{abstract}
    We generalized the conventional concept of q-plate, allowing in its definition non linear functions of the azimuthal coordinate, and simulated the resulting fields of applying this kind of element to uniformly polarized input beams, both in the near (Fresnel diffraction) and the far field (Fraunhofer diffraction) approximations. In general terms, when working in the near field regime, the chosen function defines the output polarization structure for linearly polarized input beams, and the phase of the output field for circularly polarized input beams. In the far field regime, it is obtained that when there are non-linearities in the azimuthal variable, the central singularity of the polarization field of a vector beam may bifurcate in several singularities of lower topological charge, preserving the total charge. Depending on the chosen q-plate function, different particular behaviours on the output beam are observed, which offers a whole range of possibilities for creating novel vector beams, as well as polarization critical points and singularity distributions.
\end{abstract}
 % Polarization
 % Fourier optics
 % Wave propagation
 % Optical processors, correlators, and modulators

\maketitle

\section{\label{sec:intro} Introduction}

Vector beams are known for showing a non uniform distribution of the state of polarization (SoP) \cite{zhan}. The most common beams of this kind are radially and azimuthally polarized beams, particular cases of cylindrical vector beams \cite{zhan2}, in which the SoP varies linearly with the azimuthal coordinate $\theta$. Vector beams have been widely studied because of their tight focusing properties \cite{quabis}. Besides they have potential application to communications \cite{cheng}, optical tweezers and particle micro-manipulation \cite{woerde,zhan3,kawa,zhao}, material processing \cite{duoca}, quantum entanglement \cite{gabriel} and more.

There are many methods for creating vector beams, generally divided in two categories, active and passive. Active methods consist in modifying the resonant cavity of a laser for obtaining an output vector beam, while passive ones aim to modulate the wave front of a conventional laser beam with suitable optical elements \cite{zhan}. In the latter case q-plates have become a convenient choice \cite{cardano}.

A conventional q-plate works as a half wave plate in which the director axis rotates as a linear function $q\theta$ of the azimuthal coordinate $\theta$ \cite{marrucci,marrucci2,marrucci3}. Its matrix representation in the Jones formalism has the form 
\begin{align}
    M_q(\theta) =
    \left( \begin{array}{cc}
    \cos(2q\theta) & \sin(2q\theta) \\
    \sin(2q\theta) & -\cos(2q\theta) \end{array} \right) .
\end{align}

When a linearly polarized beam passes through such an element, it becomes a vector beam with a structured polarization pattern in which the azimuth of the polarization ellipses varies as a function $2q\theta$, reaching a total rotation of $4q\pi$. On the other hand, when impinging with a circularly polarized beam, the spin to orbital conversion (STOC) phenomenon takes place: the orbital angular momentum (OAM) of the beam varies according to $\Delta l = \pm 2q$, where the plus sign applies when impinging with left circular polarization and the minus sign when impinging with right circular polarization. In other words, if an uniform left circularly polarized beam (total angular momentum +1) passes through a q-plate, it becomes an uniform right circularly polarized beam with an OAM charge $l = 2q$ (total angular momentum variation of $2(q-1)$).

This way, q-plates are high versatile elements, with many potential applications in the field of singular optics, since they allow alternately the creation of phase singularities (vortex beams with OAM) and polarization singularities (vector beams). Additionally, the possibility of using spatial light modulators (SLM), like liquid crystal displays which allow pixel to pixel phase only modulation \cite{osten,moreno16}, for the implementation of these devices, gives great flexibility for designing novel vector beams. It easily allows extending the concept of q-plate, including modulations of the polarization field that are not necessary linear in $\theta$.

Recently there have been advances towards this direction, creating q-plates with different $q$ values depending on the region of the element \cite{ji}, with non linear functions of $\theta$ for binary codification \cite{holland}, or with radial dependence for creating high order Laguerre-Gaussian beams \cite{rafa}. In this paper we propose the simulation of an element that rises from generalizing the concept of q-plate, allowing arbitrary (not necessary linear) modulations of the polarization field of a beam, in such a way that we are able to explore complex vector beams, with novel polarization structures and singularity distributions. The Jones matrix that describes this generalized q-plate (Gq-plate) is
\begin{align}
    M_{\Phi}(\theta) =
    \left( \begin{array}{cc}
    \cos(2\Phi(\theta)) & \sin(2\Phi(\theta)) \\
    \sin(2\Phi(\theta)) & -\cos(2\Phi(\theta)) \end{array} \right),
    \label{eq:q-general}
\end{align}
and it represents a half wave plate in which the director axis angle is an arbitrary function $\Phi(\theta)$. The only requirement we impose is that $\Phi$ is a continuous periodic function in $\theta$, with a period $\tau = 2\pi/n$, being $n$ any integer number.

In section \ref{sec:nlqplates} we simulate Gq-plates with non linear dependence in the azimuthal coordinate, showing its effect on the intensity, phase and polarization distributions of an uniform circular section input beam (top hat beam), for different polarizations. We show the resulting fields both in the near and the far field approximations and give an explanation of these results based on Fourier analysis. We outline in section \ref{sec:exp} a proposal for an experimental implementation using a reflective liquid crystal display (LCoS) with phase-only modulation. The main conclusions are given in section \ref{sec:conclus}.

\begin{widetext}
\begin{center}
\begin{figure}[H]
\centering
\includegraphics[width=0.748\columnwidth]{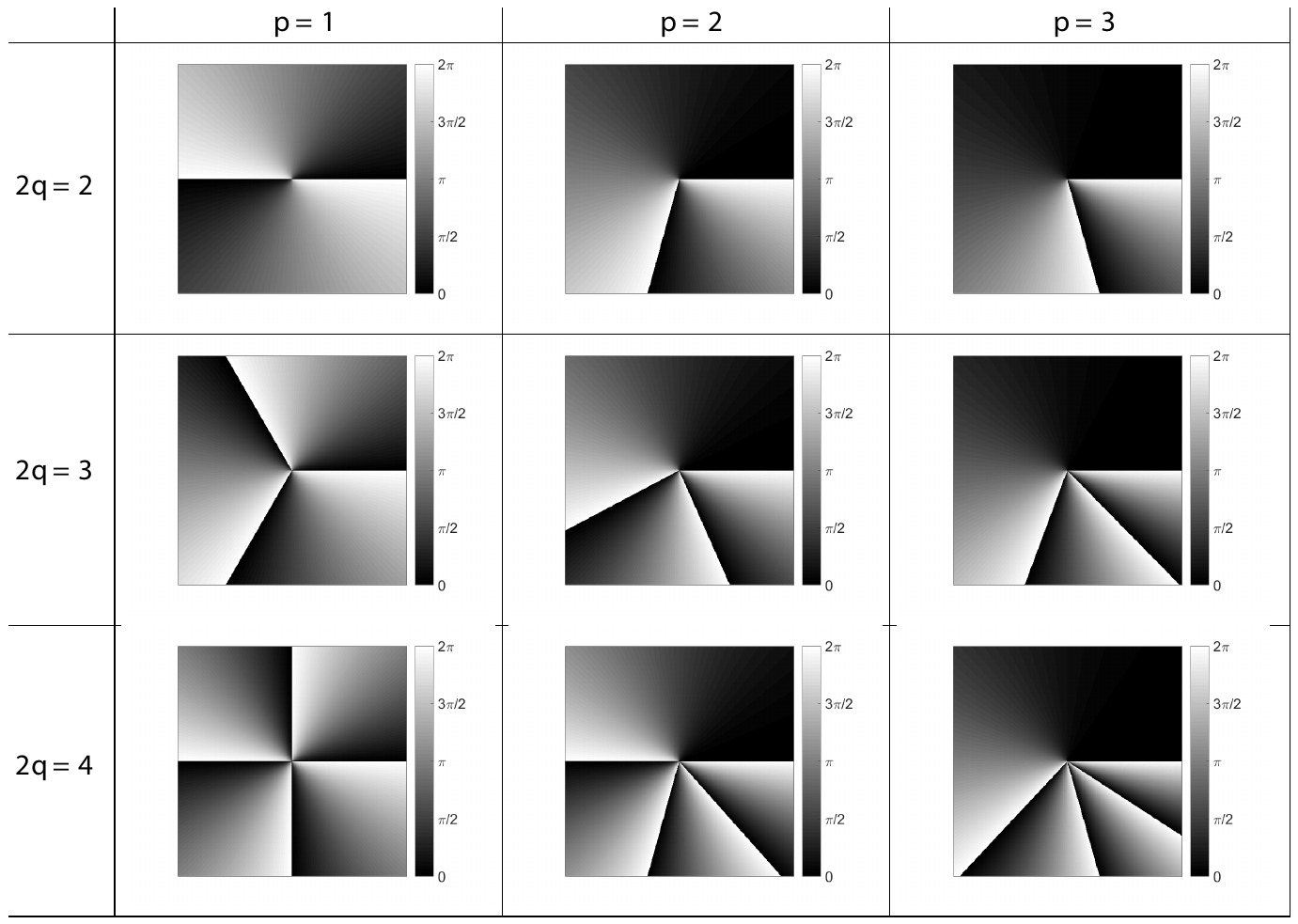} 
\caption{\small Argument function of the Gq-plates determined by a polynomial growth in $\theta$.}
\label{fig:qplates}
\end{figure}
\end{center}
\end{widetext}

\section{\label{sec:nlqplates}Non linear Gq-plates in the azimuthal variable}

When function $\Phi$ grows non-linearly in the azimuthal coordinate $\theta$, a variety of interesting effects can be observed over the resulting field's amplitude, phase and polarization structure, in this section we show some examples of the different behaviours found.

\subsection{\label{sec:poly}Polynomial growth}

We have simulated a plate defined by the non linear function $\Phi(\theta) = q(2\pi)^{(1-p)}\theta^p$. The multiplicative constant $(2\pi)^{(1-p)}$ is due to the continuity condition: $\Phi(2\pi) = q2\pi$, meaning that the total azimuthal variation is $q$ times $2\pi$, without discontinuous steps in $\Phi$ after a $2\pi$ period in $\theta$. Figure \ref{fig:qplates} shows some examples of the simulated argument function of these plates ($2\Phi(\theta)$) for different powers $p$ and values of $q$. The total azimuthal variation of the argument function is given by $2q$ times $2\pi$, which defines the topological charge of the created beams, as will be seen soon.

We studied how these Gq-plates affect an input beam with uniform phase and intensity within a circular profile (top-hat beam), simulating both the obtained field just after passing through the plate and that obtained in the far field approximation, i.e., the Fraunhofer diffraction of the former. For a vector field, this is performed simply by computing the Fourier transform of each of the $\hat{x}$ and $\hat{y}$ components of the field \cite{moreno04},

\begin{align}
\begin{split}
    \mathcal{F}\left\lbrace \mathbf{E}(x,y) \right\rbrace &= \mathbf{\tilde{E}}(u,v) \\&=
    \left( \begin{array}{c}
    \tilde{E}_s(u,v) \\
    \tilde{E}_p(u,v) \end{array} \right) \\&=
    \left( \begin{array}{c}
    \mathcal{F}\left\lbrace E_s(x,y) \right\rbrace \\
    \mathcal{F}\left\lbrace E_p(x,y) \right\rbrace \end{array} \right).
    \end{split}
\label{eq:fouriervec}
\end{align}

This is implemented numerically by means of the two dimensional discrete Fourier transform of an N $\times$ N matrix, where each element is the corresponding value of the electric field $\mathbf{E}(x,y)$ after passing through the Gq-plate.

Figure \ref{fig:CCpV} shows the intensity and polarization distribution, as well as the azimuth of the polarization ellipses, at the output plane of the Gq-plate (the last one is represented with a gray scale from $-\frac{\pi}{2}$ to $\frac{\pi}{2}$), for powers $p=1$ and $p=2$, and different values of $q$, when the input beam is linearly polarized in the vertical direction. The azimuth of the polarization ellipses is obtained by computing the phase of the complex Stokes field $S_{12}(x,y) = S_1(x,y) + iS_2(x,y)$, where $S_i(x,y)$ are the Stokes parameters of the electric field \cite{freund}.

\begin{widetext}
\begin{center}
\begin{figure}[H]
\centering
\includegraphics[width=0.9875\columnwidth]{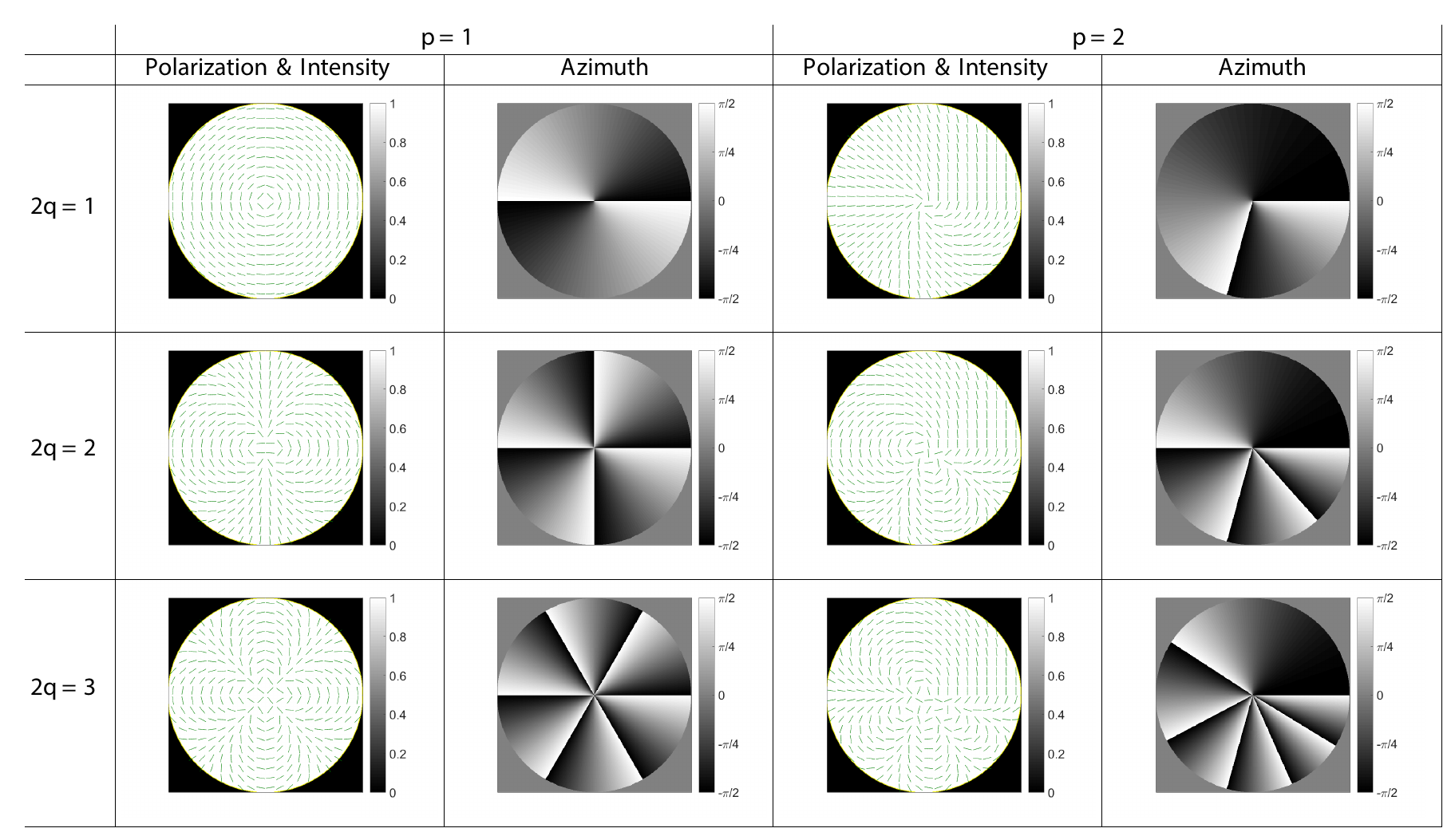}
\caption{\small (Color online) Polarization ellipses and its azimuth resulting from input vertical polarization for polynomial Gq-plates with $p=1$ and $p=2$, and different values of $q$. The output consists on vector beams in which the azimuth grows accordingly to the power $p$.}
\label{fig:CCpV}
\end{figure}
\end{center}
\end{widetext}

In the case when $p=1$ (linear q-plate) and  $q = \frac{1}{2}$ it is obtained an azimuthally polarized vector beam, whose topological charge is determined by the number of times that the polarization vector gives a complete turn around the beam axis \cite{moreno16} (in this case $2q=1$). When increasing the power to $p=2$ the topological charge does not change, while the cylindrical symmetry is lost, since the azimuth grows quadratically with $\theta$. Same behaviour is seen for higher topological charges.

In Fig. \ref{fig:CCpL} it is analyzed the behaviour of these elements when they are illuminated with circularly polarized light. In this case the polarization ellipse fields and the beam phase distributions are shown. We choose (from now on) to represent these magnitudes because, as a general fact for q-plates, when input polarization is linear, the modulation occurs in the polarization field, but when input polarization is circular, it occurs in the phase of the field, while the polarization remains uniform, changing sign 
\begin{widetext}
\begin{center}
\begin{figure}[H]
\centering
\includegraphics[width=0.9813\columnwidth]{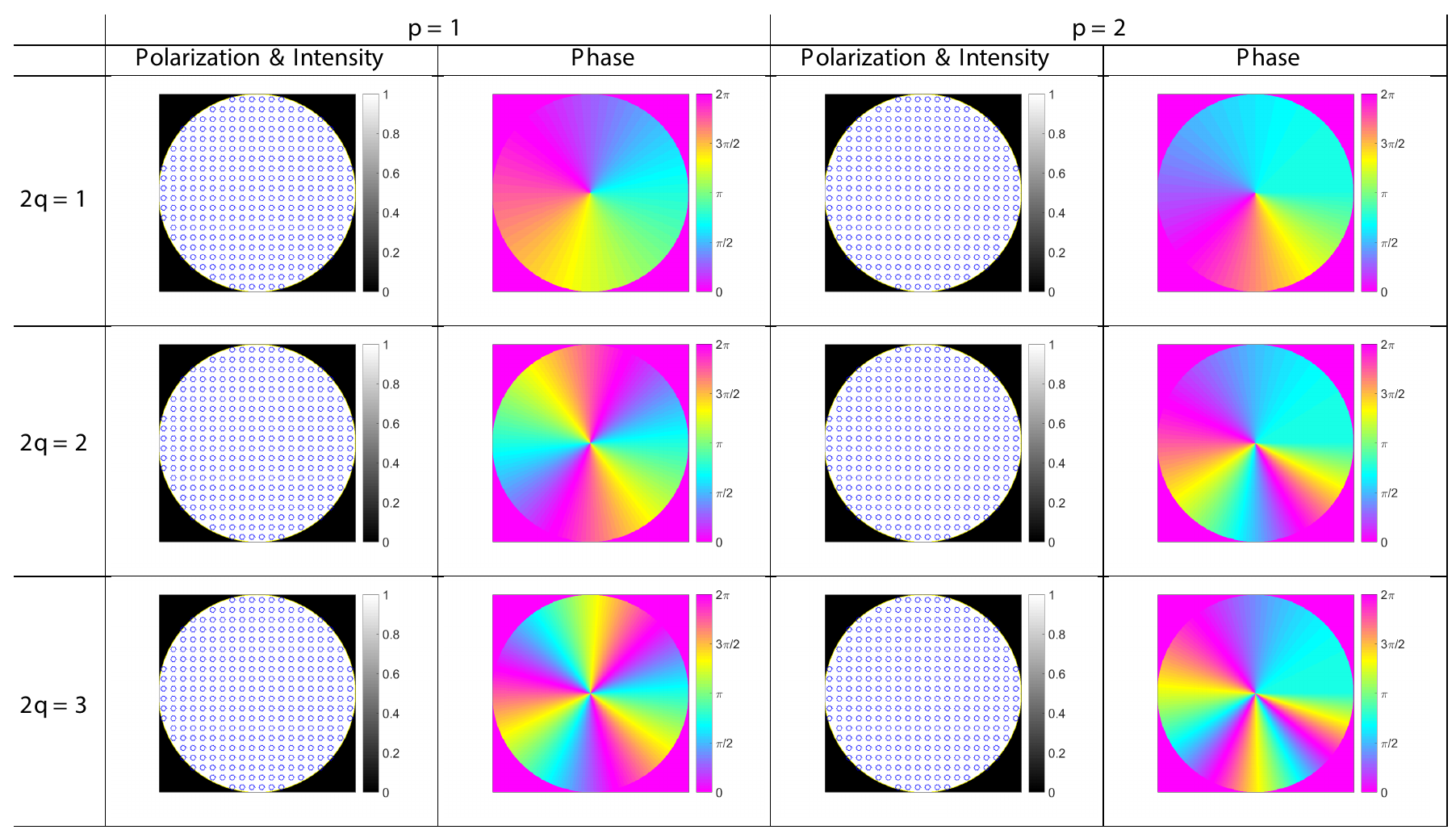}
\caption{\small (Color online) Polarization ellipses and phase distributions resulting from input left circular polarization for polynomial Gq-plates with $p=1$ and $p=2$, and different values of $q$. The output consists on uniformly right circular polarized beams in which phase grows accordingly to the power $p$.}
\label{fig:CCpL}
\end{figure}
\end{center}
\end{widetext}
due to the STOC phenomenon.

As it was previously said, for input left circular polarization it is observed the STOC phenomenon, giving place to an output uniform right circularly polarized beam, carrying OAM with topological charge $l=2q$. This is seen in the $2\pi$ turn of the beam phase around the propagation axis. The way phase grows depends on the power $p$. 

An interesting effect arises when the Fraunhofer diffraction patterns of these beams are calculated. Figure \ref{fig:CLpV} shows the result from propagating the output beams obtained when impinging with linear vertical polarization (Fig. \ref{fig:CCpV}). Both, the polarization ellipses superposed to the intensity distribution and the azimuth of the ellipses are represented. In order to plot the polarization ellipses we used a color code based on the respective form factor $f = \frac{b}{a}$, the ratio between the minor $b$ and mayor $a$ axis of the ellipse, whose sign depends on the vector sense of rotation (negative for right-handed, and positive for left-handed). There is a neighborhood around $f=0$ for which we considered polarization to be linear (green), and a neighborhood around $f = \pm1$ for which we considered polarization to be circular (blue), in any other case the polarization is elliptical (red).

For linear q-plates ($p=1$) the polarization field in the Fraunhofer regime is the same as in the q-plate plane, with a ``donut" intensity distribution, due to the central polarization vortex \cite{cardano}. Conversely, the beams obtained from non linear Gq-plates do not preserve their polarization fields. It can be seen that instead of showing a central singularity with topological charge $2q$, they show $4q$ isolated singularities each one with topological charge $\frac{1}{2}$. Furthermore, the nature of these singularities differs from the original. In Fig. \ref{fig:CLpVzoom} this fact is shown in detail for one of these singularities. It can be seen that the azimuth is not defined in these points although the form factor is, and is maximum (left circular polarization). These singularities are generally known as C-points: isolated points of circular polarization around which polarization azimuth rotates in $m2\pi$. The topological charge in this cases is $m = \frac{1}{2}$.

\begin{widetext}
\begin{center}
\begin{figure}[H]
\centering
\includegraphics[width=0.9875\columnwidth]{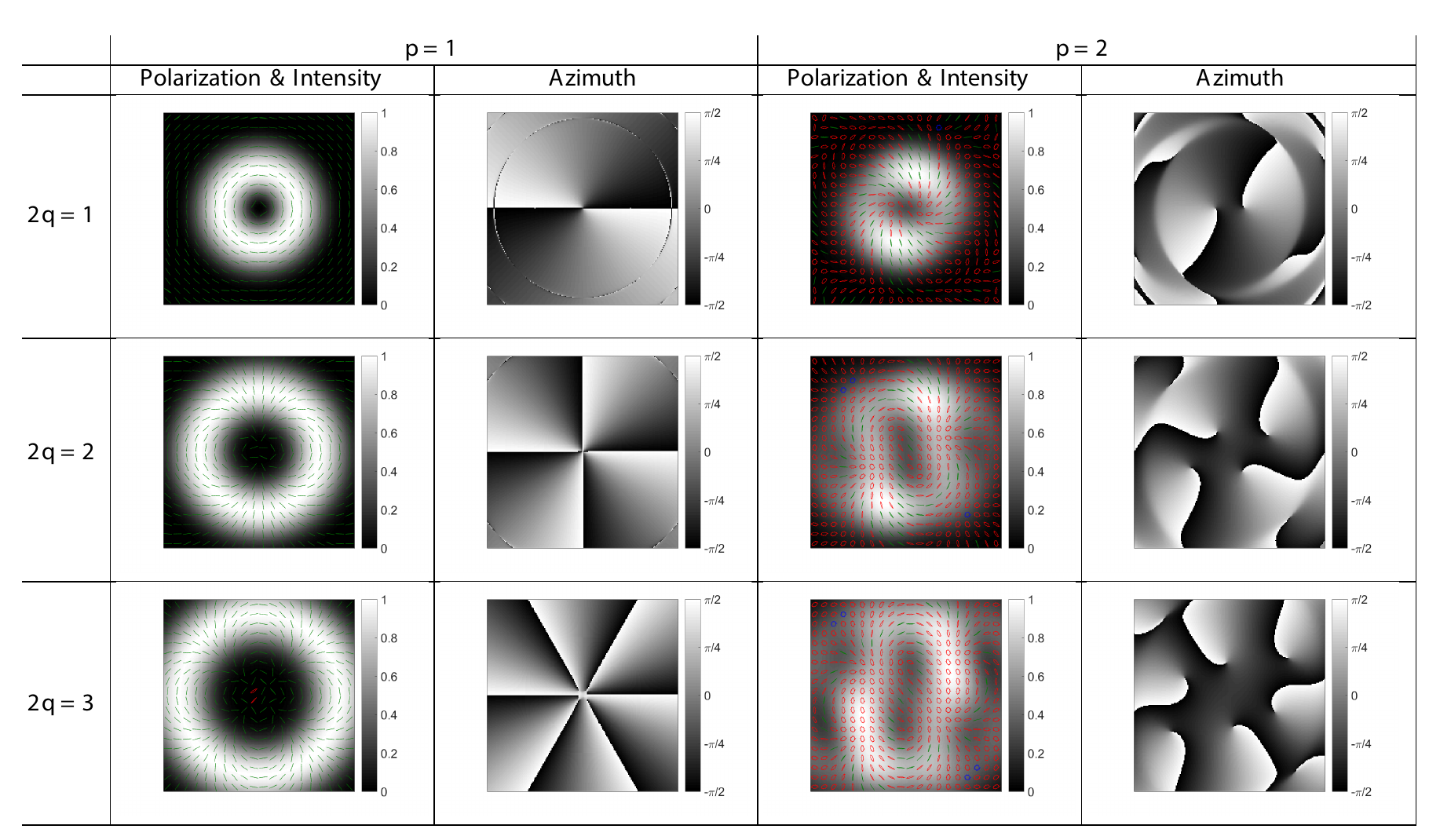}
\caption{\small (Color online) Polarization ellipses and its azimuth resulting from input vertical polarization for polynomial Gq-plates with $p=1$ and $p=2$, and different values of $q$, in the far field regime.}
\label{fig:CLpV}
\end{figure}

\begin{figure}[H]
\centering
\includegraphics[width=0.848\columnwidth]{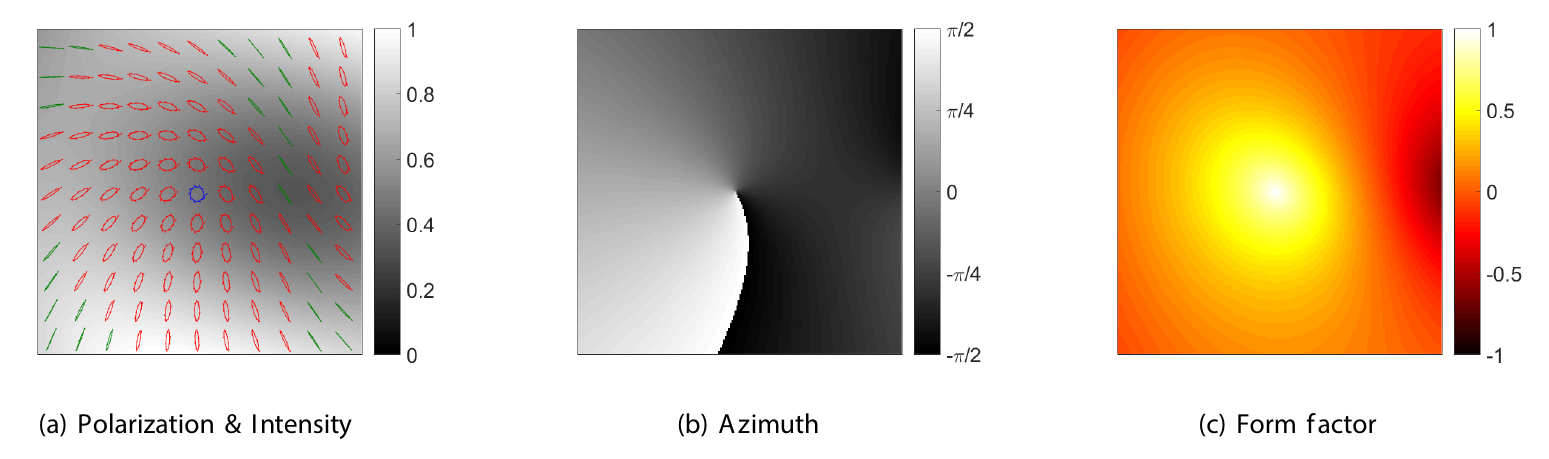}
\caption{\small (Color online) Zoom of one of the polarization singularities obtained in the far field for a polynomial Gq-plate with $q=\frac{1}{2}$ y $p=2$, when input polarization is vertical. In (a) the intensity and the polarization ellipses are depicted, in (b) and (c) the azimuth and form factor corresponding to those ellipses are respectively shown.}
\label{fig:CLpVzoom}
\end{figure}
\end{center}
\end{widetext}

Figure \ref{fig:CLpL} shows the Fraunhofer diffraction of the beams obtained when impinging onto the Gq-plates with left circularly polarized light. In this case, for a linear q-plate, the phase and polarization distributions are identical to those seen in the q-plate plane, with a central singularity (phase vortex) with topological charge $2q$, due to the creation of OAM. In the non linear case, $2q$ isolated vortexes with topological charge $1$ appear, occupying the same position as half of the C-points obtained in the case of linearly polarized light (Fig. \ref{fig:CLpV}). If impinging with right circularly polarized light (not shown), there would be another $2q$ vortexes, located according to the other half of the C-points.    

\begin{widetext}
\begin{center}
\begin{figure}[H]
\centering
\includegraphics[width=\columnwidth]{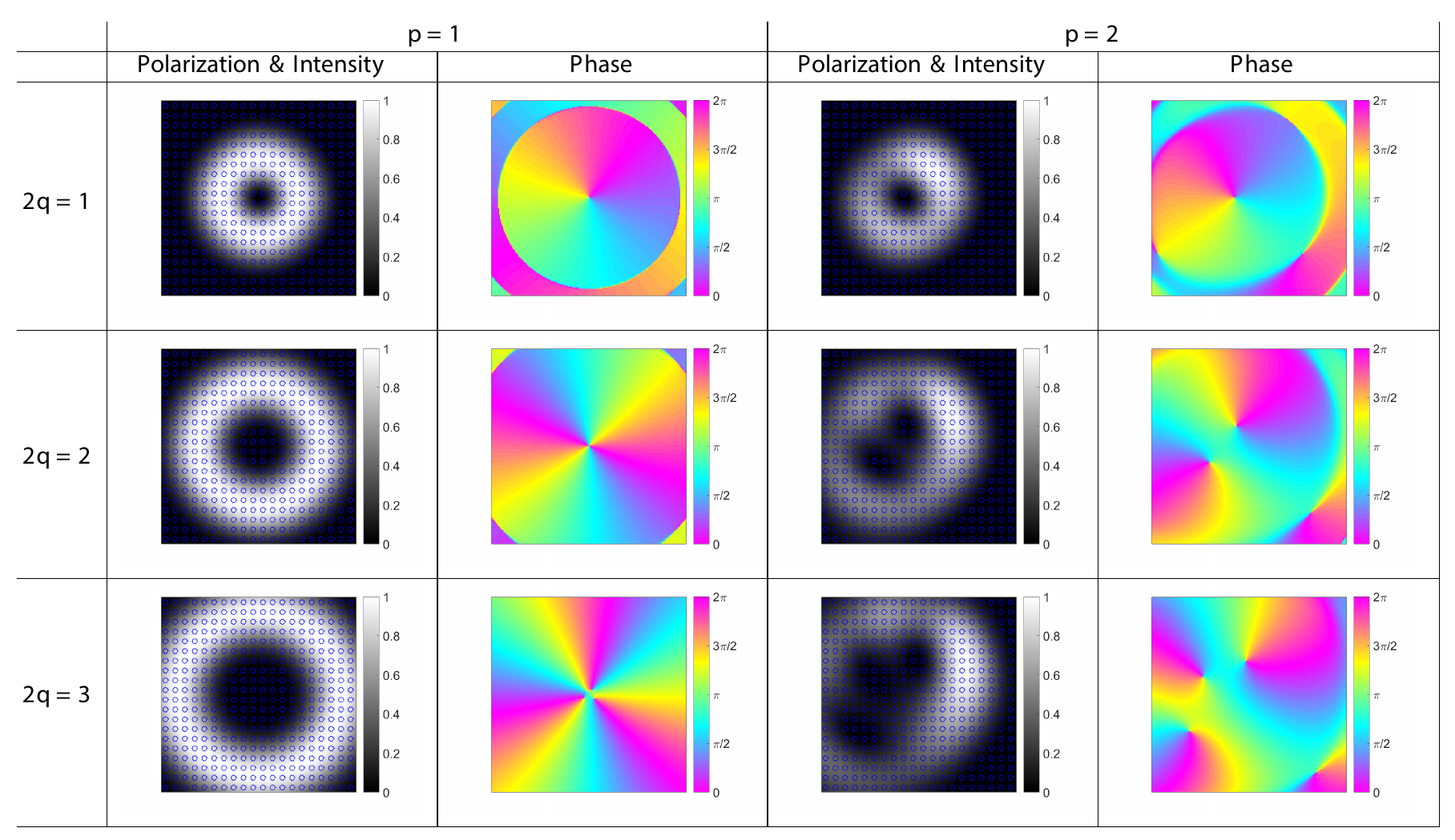}
\caption{\small (Color online) Polarization ellipses and phase distribution resulting from an input beam with left circular polarization for polynomial Gq-plates with $p=1$ and $p=2$, and different values of $q$, in the far field regime. Polarization is in every case uniform right circular.}
\label{fig:CLpL}
\end{figure}
\end{center}
\end{widetext}

This singularity splitting can be explained in terms of the far field (Fraunhofer) diffraction phenomenon. Fraunhofer field coincides with the Fourier transform of the field at the Gq-plate plane, which (for a function $g(r,\theta)$ separable in polar coordinates) can be written in terms of an infinite sum of weighted Hankel transforms \cite{goodman},
\begin{align}
    \mathcal{F}\left\lbrace g(r,\theta)\right\rbrace = \sum_{k=-\infty}^{\infty} c_k (-i)^k e^{ik\phi} \mathcal{H}_k \left\lbrace g_R (r)\right\rbrace,
\label{eq:hankel}
\end{align}
where
\begin{align}
    c_k = \frac{1}{2\pi} \int_{0}^{2\pi} g_{\Theta} (\theta) e^{-ik\theta} d\theta
\end{align}
and $\mathcal{H}_k$ is the Hankel transform operator of order $k$,
\begin{align}
    \mathcal{H}_k \left\lbrace g_R (r)\right\rbrace = 2\pi \int_{0}^{\infty} r g_{R} (r) J_k(2\pi r \rho) dr,
\end{align}
being $J_k$ the $k$th-order Bessel function of the first kind, and  $g(r,\theta) = g_R (r)g_{\Theta} (\theta)$.

With this in mind we can take as an example the cases shown in Fig. \ref{fig:CCpV} for $2q=1$ and calculate the weight distributions in each case. In the linear case ($p=1$), when input light is vertically polarized, the electric field after the q-plate is obtained from Eq. \ref{eq:q-general},
\begin{align}
\begin{split}
    \mathbf{E}_o(r,\theta) &= M_{\Phi}(r, \theta) \mathbf{E}_i(r,\theta)\\
    &= \left( \begin{array}{cc}
    \cos(\theta) & \sin(\theta) \\
    \sin(\theta) & -\cos(\theta) \end{array} \right)
    \left( \begin{array}{c}
    0 \\
    E_i   \end{array} \right) \\
    &= E_i\left( \begin{array}{c}
    \sin(\theta) \\
    -\cos(\theta)  \end{array} \right) \\ 
    &= \left( \begin{array}{c}
    E_s(r,\theta) \\
    E_p(r,\theta) \end{array} \right),
\end{split}
\end{align}
while in the non-linear case with $p=2$,
\begin{align}
    \mathbf{E}_o(r,\theta) = 
    \left( \begin{array}{c}
    E_s(r,\theta) \\
    E_p(r,\theta) \end{array} \right) = 
    E_i\left( \begin{array}{c}
    \sin(\frac{1}{2\pi}\theta^2) \\
    -\cos(\frac{1}{2\pi}\theta^2)  \end{array} \right).
\end{align}

We computed the Fourier transform of these fields according to Eqs. \ref{eq:fouriervec} and \ref{eq:hankel}, to obtain the respective weights $C_k^2 = c_{s_k}^2 + c_{p_k}^2$, where $c_s$ and $c_p$ stand for the weights of the Fourier transforms of the fields $E_s(r,\theta)$ and $E_p(r,\theta)$ respectively. Results are shown in Fig. \ref{fig:pesos}.

\begin{widetext}
\begin{center}
\begin{figure}[H]
\centering
\includegraphics[width=0.65\columnwidth]{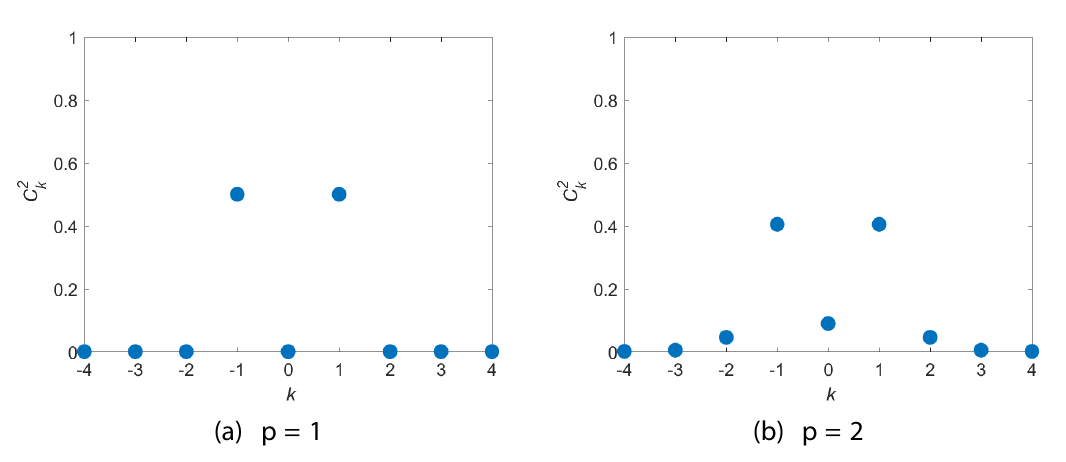}
\caption{\small Weights $C_k^2 = c_{x_k}^2 + c_{y_k}^2$ of the terms in the Fourier transforms of the fields created with a linear (a) and a non-linear (b) Gq-plate when input beam is vertically polarized.}
\label{fig:pesos}
\end{figure}

\begin{figure}[H]
\centering
\includegraphics[width=0.848\columnwidth]{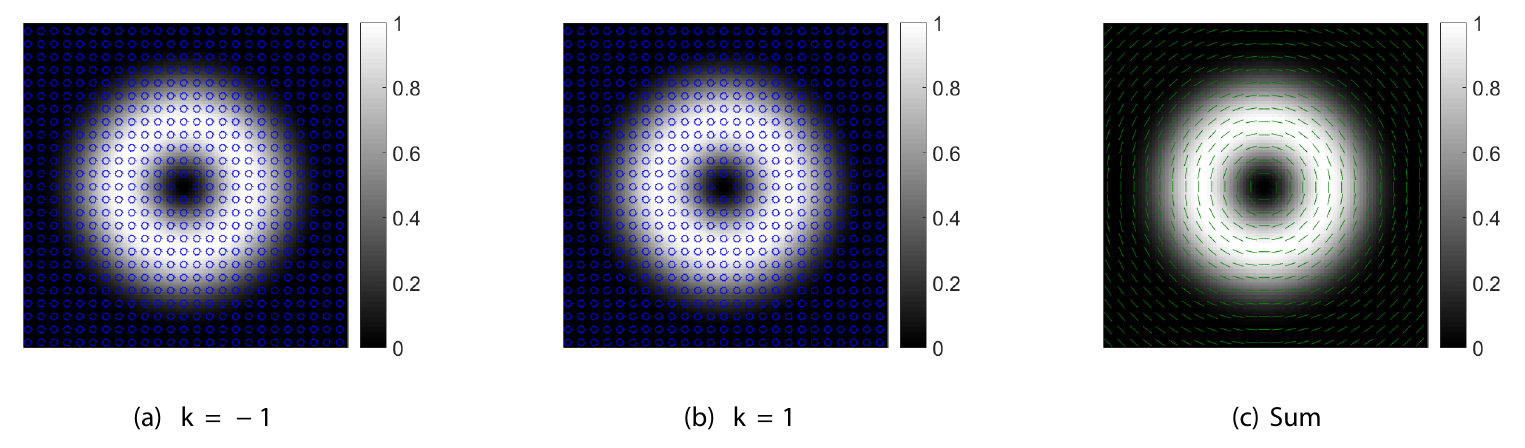}
\caption{\small (Color online) Terms with (a) $k=-1$ and (b) $k=1$ of the Fourier transform of a field created with Gq-plate with $q=\frac{1}{2}$ and $p=1$ for vertically polarized input light. (c) Shows the sum of them.}
\label{fig:term_lineal}
\end{figure}

\begin{figure}[H]
\centering
\includegraphics[width=0.848\columnwidth]{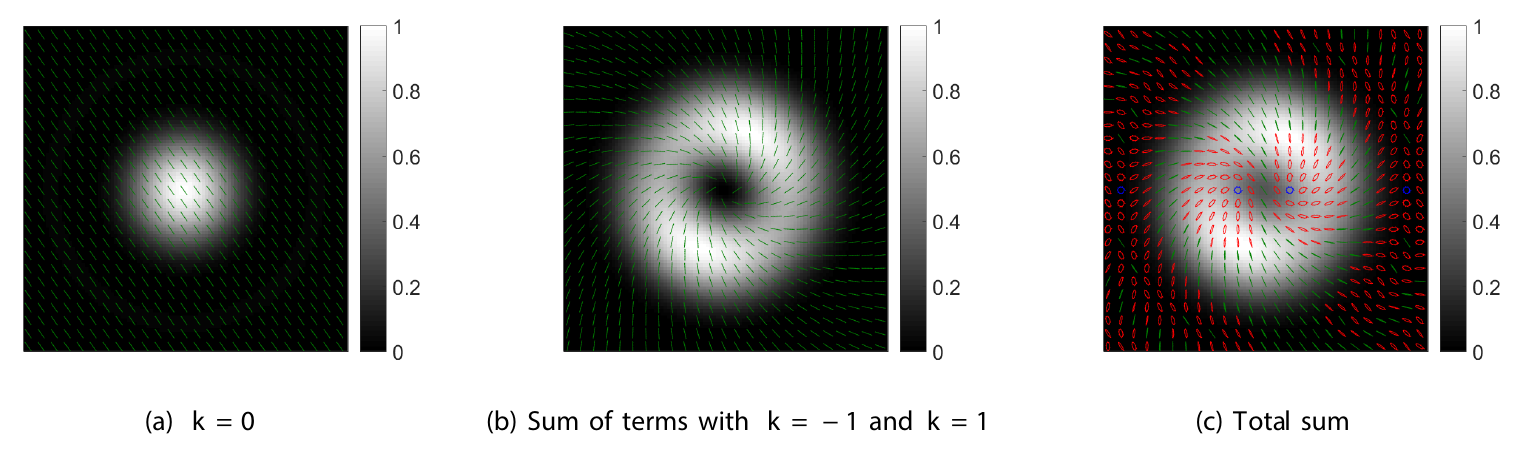}
\caption{\small (Color online) Term with (a) $k=0$, and (b) the sum of terms with $k=-1$ and $k=1$ of the Fourier transform of a field created with a Gq-plate with $q=\frac{1}{2}$ and $p=2$ for vertically polarized input light. (c) Shows the total sum.}
\label{fig:term_nl}
\end{figure}
\end{center}
\end{widetext}

In the case $p=1$ the only non-zero terms are $k=1$ and $k=-1$, both terms and the sum of them are shown in Fig. \ref{fig:term_lineal}. This, as expected, is identical to the field shown in Fig.\ref{fig:CLpV}. On the other hand, in the case with $p=2$, while terms $k=1$ and $k=-1$ remain the most significant, other terms arise, in particular that with $k=0$. This term contributes with a Bessel function $J_0$, giving non-zero intensity at the propagation axis, and hence
\begin{widetext}
\begin{center}
\begin{figure}[H]
\centering
\includegraphics[width=0.5\columnwidth]{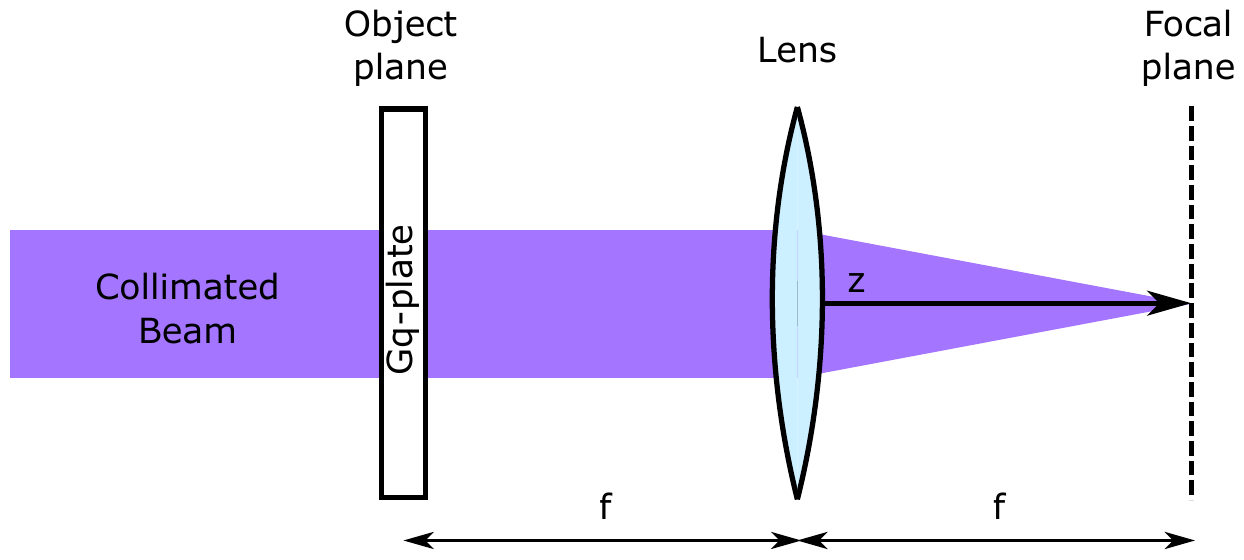}
\caption{\small (Color online) Scheme of the optical device used for computing the beam evolution as it propagates from the Gq-plate to the far field regime.}
\label{fig:fourier}
\end{figure}
\begin{figure}[H]
\centering
\includegraphics[width=0.7998\columnwidth]{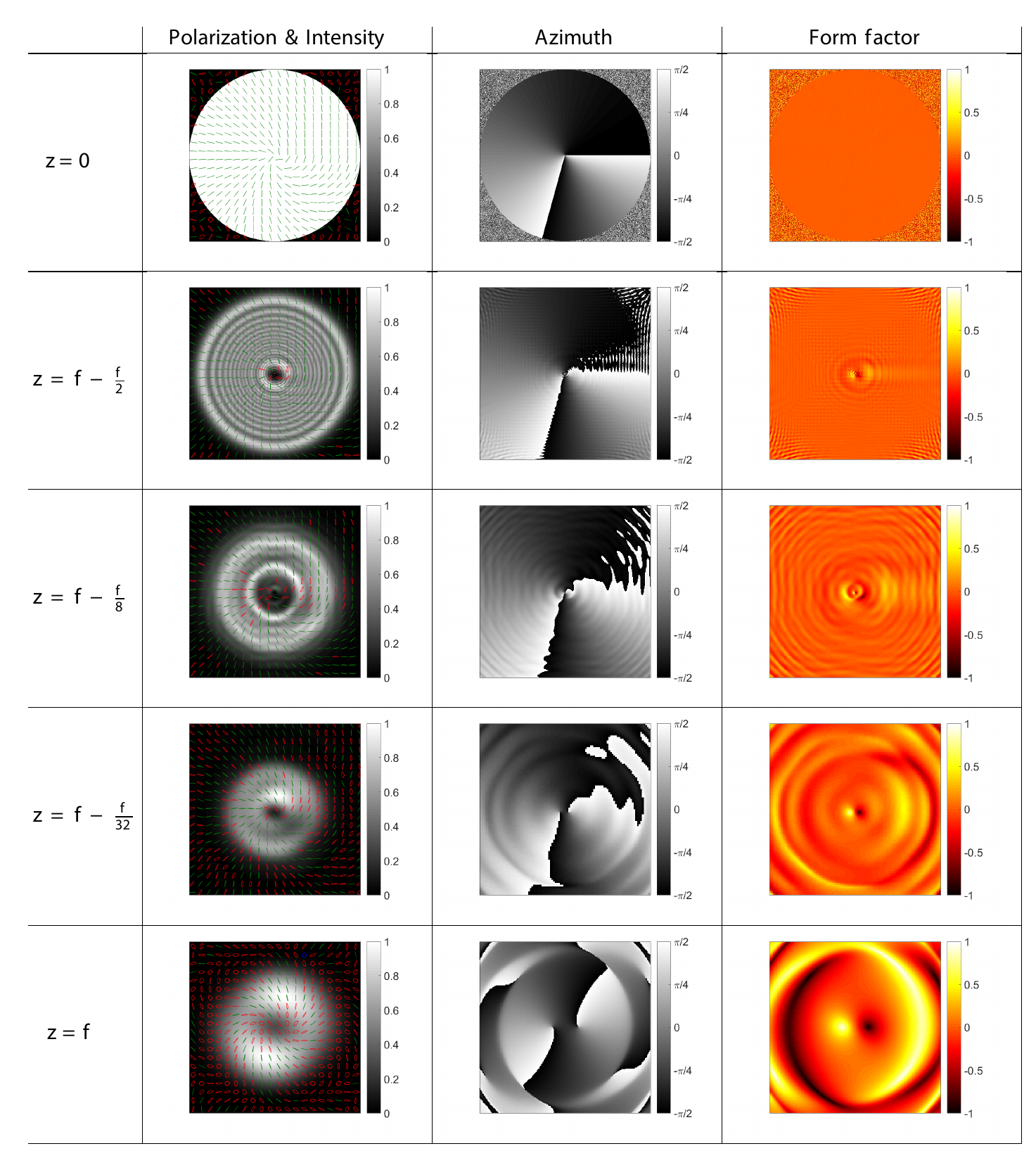}
\caption{\small (Color online) Fresnel diffraction pattern at different planes resulting from a polynomial Gq-plate with $q=\frac{1}{2}$ y $p=2$, when the input light is vertically polarized. The intensity distribution superposed to the polarization ellipses are shown in the first column. The azimuth and form factor are depicted in the second and third column respectively.}
\label{fig:FpV}
\end{figure}
\end{center}
\end{widetext}
destroying the dark singularity, as shown in Fig. \ref{fig:term_nl}. Since at the beam axis the only non-zero term is the $k=0$, the polarization there is linear at $-55^\circ$. In the outer region, the predominant terms $k=1$ and $k=-1$ create a cylindrically polarized vector beam. The continuous transition between the SoP at the center and in the outer region, and the phase difference between these terms due to the factor $(-i)^k$ in Ec. \ref{eq:hankel}, force the apparition of two C-points.

It is interesting to analyze how these fields evolve from their pass through the Gq-plate to the far field regime, and how is the transition between the central singularity and the multiple isolated singularities. For that purpose we have simulated the device depicted in Fig. \ref{fig:fourier}. We have added a quadratic phase to the field, representing the effect of a lens, and numerically calculated the Fresnel diffraction integral for transverse planes at different distances $z$, from the lens plane $z=0$ to the focus $z=\mathbf{f}$. The field distribution at the focal plane is equal to that obtained by directly Fourier transforming (Fraunhofer diffraction). Intermediate planes show the beam polarization structure in the near field regime.

Figure \ref{fig:FpV} shows some of the propagated fields for the case $q=\frac{1}{2}$, $p=2$, when the input beam is verically polarized. A short movie showing the complete evolution is included in the Supplemental Material \cite{supplemental}. For distances close to the lens plane, i.e., when $z < \mathbf{f} - \mathbf{f}/2$, the intensity of the beam remains approximately constant, and the polarization structure is the same as in the Gq-plate plane, with a central singularity, this is in agreement with the results reported in reference \cite{moreno16}. For $z = \mathbf{f} - \mathbf{f}/8$ the polarization field begins to distort, showing regions in which polarization is elliptical. Further on, e.g., when $z = \mathbf{f} - \mathbf{f}/32$, two critical points of the form factor clearly appear, in the center of the beam, and the central singularity is divided in two, although most of the polarization structure remains similar to that at $z=0$. Finally at the focal plane the polarization structure is totally distorted, showing elliptic polarization around two C-points with topological charge $\frac{1}{2}$, and no central point with null intensity. It's remarkable that the field, which in principle does not show any symmetry, in the far field regime gains symmetry with respect to the transformation that rotates the beam $\pi$ radians around its axis and inverts the rotation of the polarization vector. When input polarization is left (right) circular, polarization turns right (left) circular and remains uniform during propagation. Phase distribution remains unmodified for distances close to the lens plane, and then distorts into the singularities discussed earlier.

\subsection{\label{sec:sinu}Sinusoidal variation}

Another interesting possibility is that of designing an element that modulates the electric field without adding any net topological charge, e.g., a Gq-plate defined by an oscillating function in $\theta$. That is the case for the function $\Phi(\theta) = -\frac{\pi}{2}(\cos(q\theta)-1)$, where $2\Phi$ (Eq. \ref{eq:q-general} matrix argument) oscillates between $0$ y $2\pi$, being $q$ the number of periods for $\theta\in[0,2\pi]$. Figure \ref{fig:qplates_sin} shows some examples of this function for different values of $q$.

\begin{widetext}
\begin{center}
\begin{figure}[H]
\centering
\includegraphics[width=0.48\columnwidth]{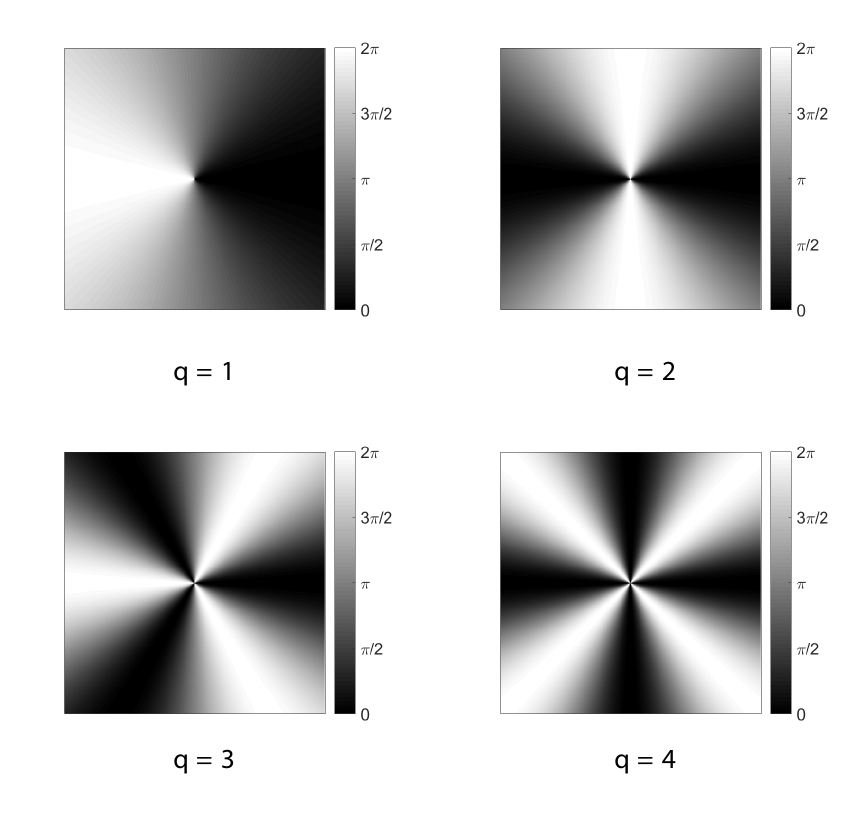}
\caption{\small Argument function ($2\Phi(r,\theta)$) for sinusoidal Gq-plates.}
\label{fig:qplates_sin}
\end{figure}

\begin{figure}[H]
\centering
\includegraphics[width=0.995\columnwidth]{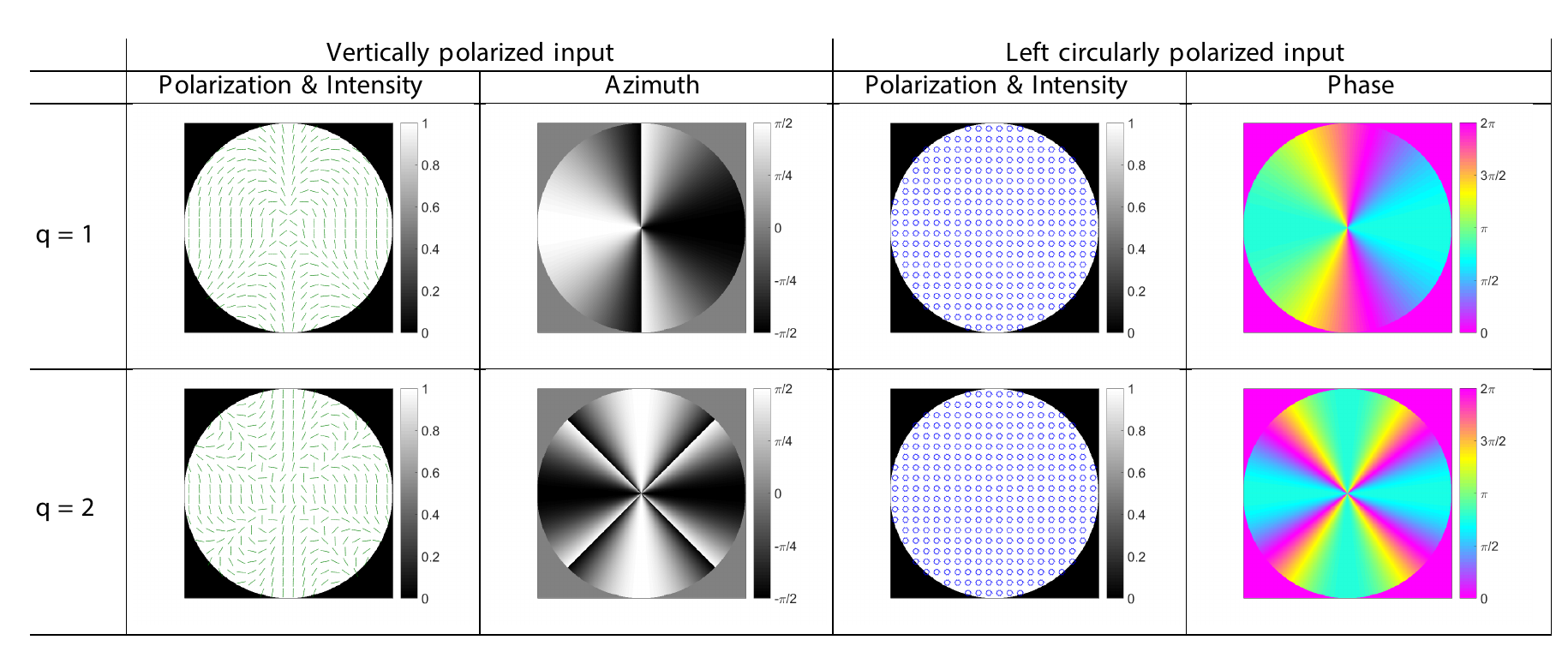}
\caption{\small (Color online) Polarization distributions, azimuth and form factor provided by sinusoidal Gq-plates with $q=1$ and $q=2$, for vertical and left circular input polarizations.}
\label{fig:CCs}
\end{figure}

\begin{figure}[H]
\centering
\includegraphics[width=0.768\columnwidth]{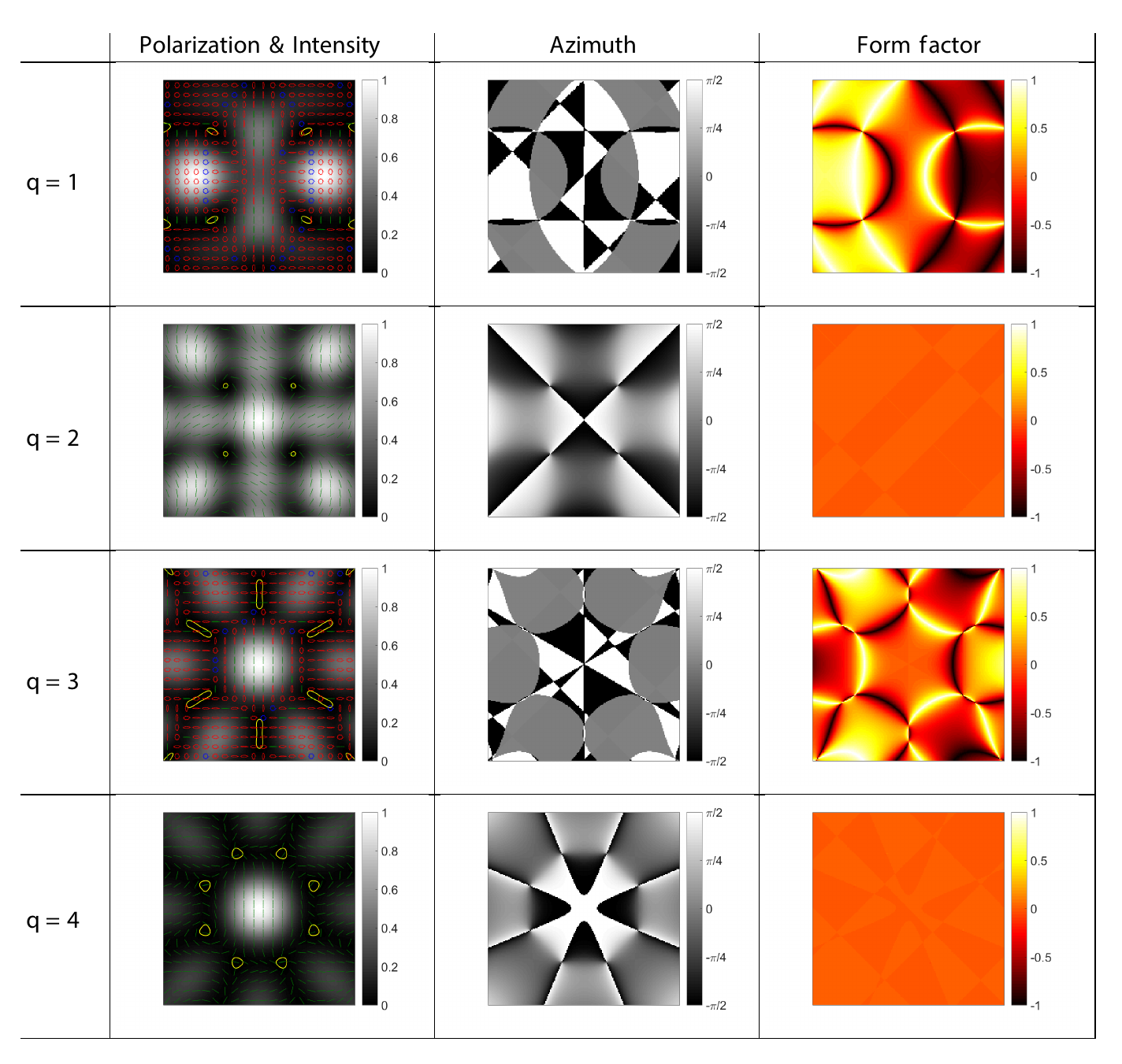}
\caption{\small (Color online) Far field polarization distribution, azimuth and form factor obtained for different values of $q$, when input polarization is vertical. Yellow lines demarcate minimum intensity areas.}
\label{fig:CLsV}
\end{figure}
\end{center}
\end{widetext}

Figure \ref{fig:CCs} shows the results obtained from some of these Gq-plates when vertically and left circularly polarized beams are used to illuminate the element. For vertically polarized input, it is achieved an output with uniform phase distribution and a polarization whose azimuth oscillates, while for circularly polarized input an oscillating phase distribution with uniform polarization is obtained. Again, as discussed in the previous section, at the exit of the Gq-plate the function $\Phi$ defines the polarization structure when impinging with linear polarization and the phase structure when impinging with circular polarization. Novel effects occur in the far field beyond the Gq-plate, as shown in Fig. \ref{fig:CLsV}. There it is shown the Fraunhofer diffraction fields obtained after impinging onto these kind of Gq-plate with vertically polarized light. Yellow contours delimit the minimum intensity areas, defined as the regions with less than $0.5\%$ of maximum intensity.

It can be observed an interesting behaviour that depends on the parity of $q$. For odd $q$ values, the diffraction pattern shows several intensity minima, which match with corresponding saddle points in the form factor, keeping the azimuth of the polarization ellipses according to the input beam (except for $\frac{\pi}{2}$ rotations). On the other hand, for even $q$ values, form factor remains uniformly zero (linear polarization), while $2q$ dark azimuth singularities arise, which match with corresponding $2q$ intensity minima, distributed geometrically around the beam axis. These singularities have alternate $\pm1$ topological charges, adding up 0. Topological charge is measured as the times that the azimuth completes a $2\pi$ rotation along a closed path around the singularity, and the sing is provided by the sense of rotation, positive charge singularities are known as flowers, while negative charge singularities are known as webs \cite{freund}.

\begin{widetext}
\begin{center}
\begin{figure}[H]
\centering
\includegraphics[width=0.5421\columnwidth]{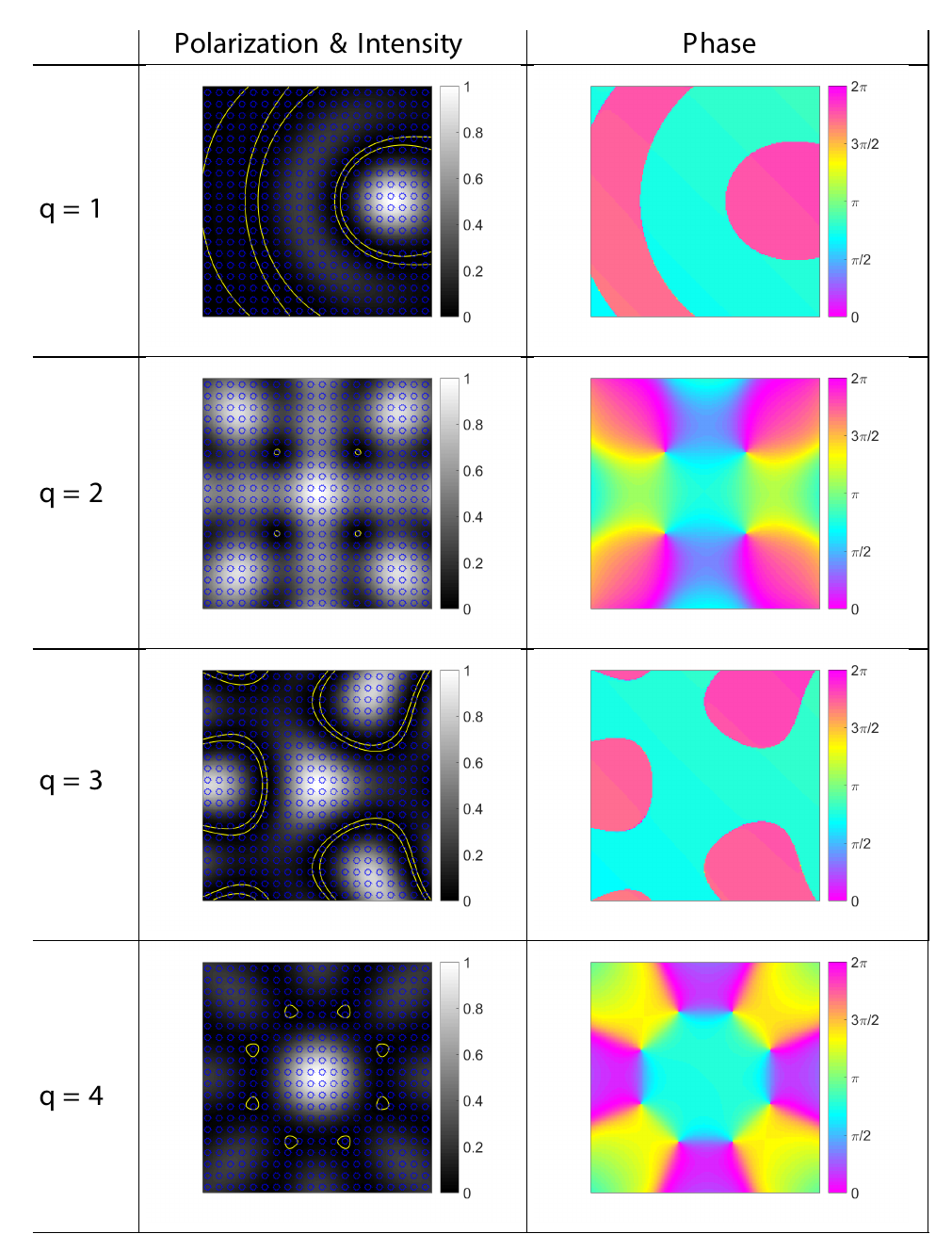}
\caption{\small (Color online) Far field intensity, polarization ellipses and phase distribution obtained for different values of $q$, when input polarization is left circular. Yellow lines demarcate minimum intensity areas.}
\label{fig:CLsL}
\end{figure}
\end{center}
\end{widetext}
This can be explained as well in terms of the Fourier transform decomposition of Ec. \ref{eq:hankel} discussed earlier. For instance, when input light is vertically polarized, the electric field at the exit of the Gq-plate is
\begin{align}
    \mathbf{E}_o(r, \theta) = 
    \left( \begin{array}{c}
    E_s(r,\theta) \\
    E_p(r,\theta) \end{array} \right) =
    E_i\left( \begin{array}{c}
    \sin(-\pi(\cos(q\theta)-1)) \\
    -\cos(-\pi(\cos(q\theta)-1))  \end{array} \right).
\end{align}
If we look at the $c_k$ values of the Fourier transform of these field, it is found that for odd $q$ values, $E_s$ shows in its expansion only terms with odd $k$ value, while $E_p$ shows only even valued terms. Since all $c_k$ values are real in this case, this means that the horizontal component of the electric field has a phase factor $(-i)^k = \pm i$, while the vertical component has a phase factor $(-i)^k = \pm 1$. Then, the phase difference between these components has to be $\pm \frac{\pi}{2}$, giving polarization ellipses vertically or horizontally oriented, with a form factor depending on the amplitude ratio. On the other hand, for even $q$ values, Fourier expansion of both components of the electric field shows only terms with even $k$ values, so phase difference between them must be $0$ or $\pm \pi$, giving now linear polarization with azimuth depending on the amplitude ratio.

Similar distinction occurs when the polarization of the input beam is left circular, as shown in Fig. \ref{fig:CLsL}. For odd $q$ values, there are no isolated singularities, but minima valleys which set a $\pi$ step in the beam phase. Regarding Fig. \ref{fig:CLsV} it can be seen that these minima valleys match left circularly polarized regions resulting from vertically polarized input. A vertically polarized beam can be described as the balanced superposition of left and right circularly polarized beams, and after passing through the Gq-plate, left circular polarization turns right, and vice versa. Then, its reasonable that when input light is left circularly polarized, regions of the output beam corresponding to left circular maxima show no intensity. On the other hand, for even $q$ values there are $2q$ phase vortexes carrying alternate topological charges (OAM) of $\pm1$, matching respective intensity isolated minima. Again, the total topological charge adds up to 0. Comparing with the case with linearly polarized input, intensity distributions are the same, changing polarization vortexes into phase vortexes now. This is consistent, since sinusoidal Gq-plates with even $q$ values seems to modulate left and right circularly polarized light in the same way. These kind of distribution of optical vortexes with alternate charges around the beam axis may have potential application in optical trapping and micro-manipulation \cite{gecevicius}.

\begin{widetext}
\begin{center}
\begin{figure}[H]
    \centering
    \includegraphics[width=0.5\columnwidth]{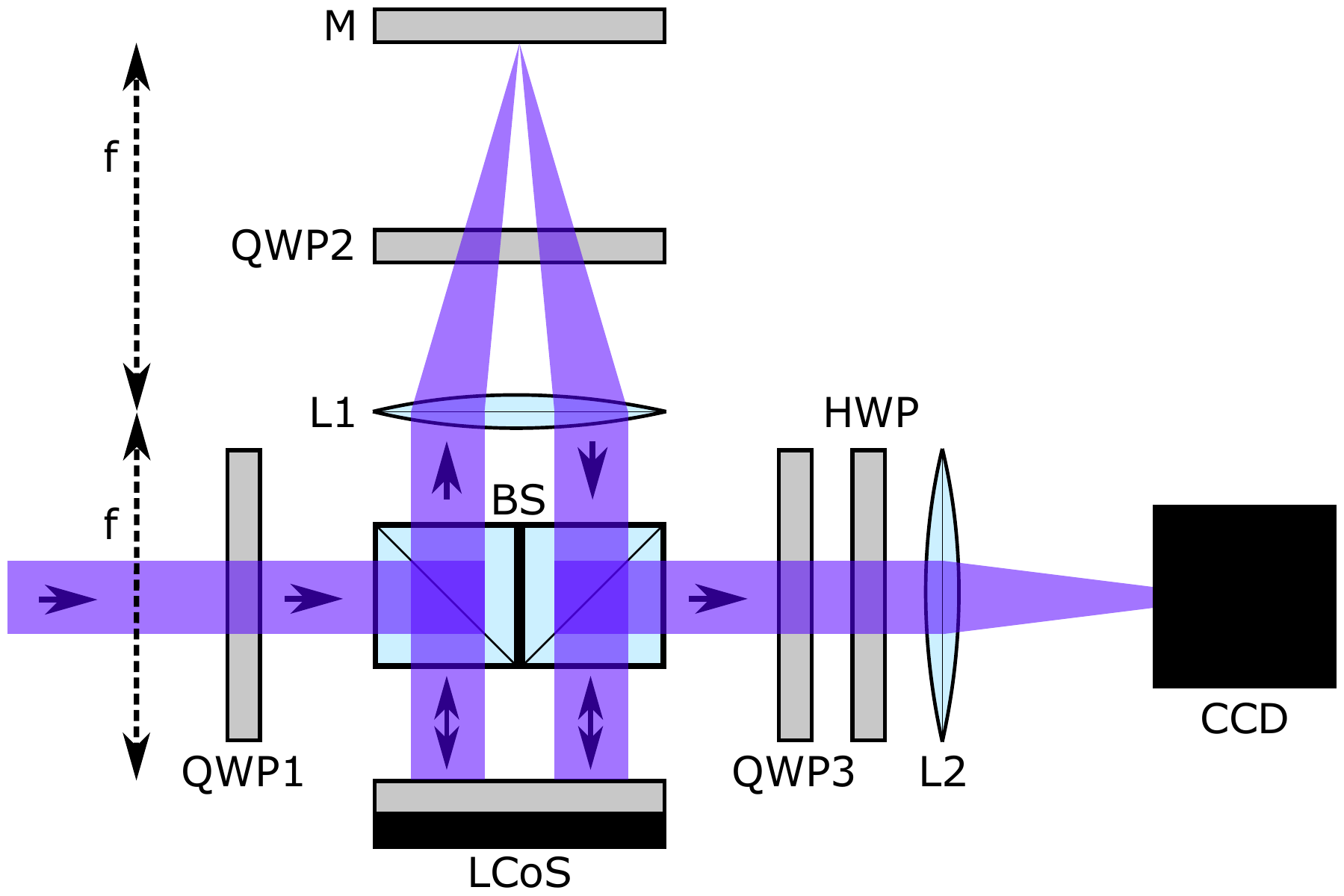}
    \caption{\small (Color online) Experimental compact device proposed for implementing generalized q-plates using a reflective PA-LCoS.}
    \label{fig:dispositivo}
\end{figure}
\end{center}
\end{widetext}

\section{\label{sec:exp}Experimental implementation proposal}

Here we propose a tentative compact device for implementing the generalized q-plates. The device is designed for modulating independently the phase of the orthogonal components of an input electric field, using a commercially available parallel aligned reflective liquid crystal on silicon (PA-LCoS) display. This kind of displays introduce a programmable phase modulation to one linear component of the field (let us suppose that the director of the LC molecules is horizontally oriented). The proposed setup is sketched in Fig. \ref{fig:dispositivo} and is based on a similar architecture used in \cite{moreno12} where the authors employ a parallel aligned transmission display that is an unusual device.

Let us describe how to get the desired modulation with this setup. The incident collimated beam is reflected by means of a first beam-splitter (BS) onto one half of the LCoS, where a phase of $\psi = 2\Phi(r,\theta)$ is added to the horizontal component of the electric field. Then the beam propagates along a 4f system, where a quarter wave plate (QWP2), oriented at $45^\circ$ respect to the LC director, rotates the polarization in $\frac{\pi}{2}$, due to the double passage. At the focus of L1 a mirror (M) redirects the rotated beam onto the other half of the LCoS, where a phase $-\psi = -2\Phi(r,\theta)$ is added to the remaining orthogonal component. By means of a second beam-splitter, the modulated beam is reflected towards a CCD. Wave plates QWP1 (oriented at $45^\circ$), QWP3 (oriented at $-45^\circ$) and HWP (oriented at $0^\circ$) are required \cite{moreno16} for the matrix representation of the whole device as that of the Gq-plate (Eq. \ref{eq:q-general}). Lens L2 is useful for measuring the output beam at different propagation distances, between near and far field regimes, it can be removed for observing directly the intensity obtained at the exit of the Gq-plate.

\section{\label{sec:conclus} Conclusions}

We simulated generalized q-plates (Gq-plates) which modulate the incident beam with non-linear functions of the azimuthal variable, and studied their effects on uniform linearly and circularly polarized beams.

In the near field approximation it is found that for linearly polarized input, the output polarization structure is given by the argument function of the Gq-plate, showing a central singularity, characteristic of conventional vector beams, while for circularly polarized input the output phase structure is the one modulated, giving place to the generation of OAM and the inversion of the polarization sense (STOC phenomenon).

In the far field regime, it is found that when losing linearity in the azimuthal variable, the conventional central singularity bifurcates in several singularities of minimum topological charge. In the cases where the input light is linearly polarized, the output beam can exhibit, either C-points with topological charge $\pm\frac{1}{2}$, as well as other types of critical points of the form factor, or dark polarization singularities (flowers/webs). Circularly polarized input beams, result in the appearance of phase vortexes, carrying OAM with topological charge $\pm1$. The intensity profiles and singularity distributions in each case depends on the particular chosen function $\Phi$, giving the chance to model distributions of any optical singularity known. This results were analyzed and discussed in terms of Fourier decomposition for separable functions in cylindrical coordinates.

Capability of representing arbitrary functions, by means of an experimental device based on a PA-LCoS, opens a wide range of possibilities for experimental creation of novel vector beams and distributions of polarization singularities or critical points of different kind, with potential use in the field of singular optics.

\begin{acknowledgements}
    This work was supported by UBACyT Grant No. 20020170100564BA, and ANPCYT Grant No. PICT 2014/2432. M.V. holds a CONICET Fellowship.
\end{acknowledgements}

%\bibliography{ref}


\end{document}